# A Standard Grammar for Temporal Logics on Finite Traces

- https://arxiv.org/abs/2012.13638
- marcofavorito/tl-grammars

Document version: v0.2.0

**WARNING**: this version v0.2.0 is a draft. You are encouraged to email the contact author for any comment or suggestion.

## Authors

- **Marco Favorito**
  0000-0001-9566-3576 · marcofavorito · https://marcofavorito.me
  Department of Computer, Control and Management Engineering - Sapienza University of Rome

## Abstract

The heterogeneity of tools that support temporal logic formulae poses several challenges in terms of interoperability. In particular, a standard syntax for temporal logic on finite traces, despite similar to the one for infinite traces, is currently missing. This document proposes a standard grammar for several temporal logic formalisms interpreted over finite traces, like Linear Temporal Logic (LTLf), Linear Dynamic Logic (LDLf), Pure-Past Linear Temporal Logic (PLTLf) and Pure-Past Linear Dynamic Logic (PLDLf).

## Introduction

This section explains the motivations behind the existence of this standard, states the goals of the standard, describes the notation conventions used thorough the document, and lists the normative references[1].

## Motivation

Temporal logics have a long history [1]. One of the most influential formalisms is Linear Temporal Logic (LTL) [2], which has been applied for program specification and verification. The variant with finite trace semantics, LTLf, has been introduced in [3].

Linear Dynamic Logic (LDL) [4;] is the extension of LTL with regular expressions (RE). The idea behind LDL is to have a formalism that merges the declarativeness and convenience of LTL, as expressive as star-free RE, with the expressive power of RE. The variant over finite traces, LDLf, has been proposed in [3]. The syntax that naturally supports empty traces has been employed in [5] for LTLf/LDLf.

Recently, a finite trace variant has been proposed also for the pure-past versions of LTLf and LDLf, namely Pure-Past Linear Temporal Logic (PLTLf) and Pure-Past Linear Dynamic Logic (PLDLf) [6].

The topic has gained more and more attention both in academia and industry, also because such logics have been considered compelling from a practical point of view. Among areas of Computer Science and Artificial Intelligence, we encounter reactive synthesis [7], model checking [8], planning with temporal goal [9], theory of Markov Decision Process with non-Markovian rewards [10], business processes specification [11], just to name a few. For what concerns industry applications, Intel proposed the industrial linear time specification language `ForSpec` [12], and the IEEE association standardized the *Property Specification Language* (PSL) [13]. Both standards witness the need of specifications based on LTL and regular expressions. Also, the research community has proposed a plethora of software tools and libraries to handle LTL and/or LDL formulas for a variety of purposes: `Spot` [14,15], `Owl` [16], `SPIN` [17] for the infinite-trace semantics, and `Syft` [18], `Lisa` [19], `FLLOAT` [20,21], `LTLf2DFA` [22], `Lydia` [23] for the finite trace setting. Another related work is represented by TLSF v1.1 [24], although its focus is on a format for LTL synthesis problems.

All these tools and formats assume the input formulae to be written in a certain grammar. Unfortunately, as often happens when dealing with parser implementations with lack of coordination, the grammars to represent the formulae have some form of discrepancies; e.g. different alternative ways to denote boolean conjunctions or temporal operators, different lexical rules to describe the allowed atomic propositions or boolean constants, underspecifications on how to handle special characters (linefeed, tab, newline, etc.), how to handle associativity of the operators.

# Goals

To enhance interoperability between the aforementioned tools, this document proposes a standard grammar for writing temporal logic formulae. In particular, we specify grammars for:

- Linear Temporal Logic on finite traces (LTLf)
- Linear Dynamic Logic on finite traces (LDLf)
- Past Linear Temporal Logic on finite traces (PLTLf)
- Past Linear Dynamic Logic on finite traces (PLDLf)

Note that, despite the syntax is very similar between the finite trace and the infinite trace variants, it is not the same for some operators. For instance, in LTL there is no *weak next* operator, wheras in LTLf it is the dual operator (under negation) of the *next* operator.

We would like this standard to be:

- An *open* standard, fostering collaboration and contributions from the research community;
- As much compliant as possible to existing and widely used tools;
- Written by researchers, for researchers. In other words, this is not strictly tight to industrial needs; for instance, we deliberately dropped the modeling of multiple clock and reset signals of `ForSpec` and `PSL`, as these are constructs not relevant for domains outside formal verification.
- Tool-agnostic. Often, grammars are reported alongside software manuals and descriptions. Instead, our aim is to propose a common denominator for all the grammars in use.

# Notation

We describe the syntax in Extended Backus-Naur Form (EBNF) [25]. We follow the notation used for the specification of XML [26]; we discarded the EBNF standard version ISO/IEC 14977 [27], as it has been often rejected by the community of those who write language specifications for a variety of reasons [28,29].

# Normative

We refer to [30] for requirement level key words. We also refer to Unicode standard [31,32] to define legal characters. For versioning this standard, we use SemVerDocs [33], inspired by SemVer [34].

# Common definitions

In this section, we describe syntactic rules shared across every logic formalism.

## Characters

Parsers MUST be able to accept sequence of *characters* (see definition below) which represent temporal logic formulae. A *character* is an atomic unit of text as specified by ISO/IEC 10646:2020 [31]. Legal characters are tab, carriage return, line feed, and the ASCII characters of Unicode and ISO/IEC 10646.

The range of characters to be supported is defined as:

```
Char ::= #x9 | #xA | #xD | [#x20-#x7e]
```

That is, the character tabulation, line feed, carriage return, and all the printable ASCII characters.

## Boolean constants

We use `true` and `false`, to denote *propositional* booleans, and `tt` and `ff`, to denote *logical* booleans. Note that `true` != `tt`, as `true` requires reading *any* symbol from the trace, e.g. in LTLf, whereas `tt` is the tautology. Similarly, `false` != `ff` as `false` requires reading no symbol, whereas `ff` is the contradiction. For `false` and `ff` the difference is a bit more blurred, but we considered it better to keep them for symmetry with the positive case.

```
True  ::= true
False ::= false
TT    ::= tt
FF    ::= ff
PropBooleans  ::= TRUE | FALSE
LogicBooleans ::= TT | FF
```

## Atomic Propositions

An atomic proposition is a string of characters. In particular, it can be:

- any string of printable characters, excepted the quotation character used (see `QuotedName`)
- any string of at least one character that starts with `[a-z_]` and continues with `[a-z0-9_]`, and that is not a reserved keyword.

Unquoted strings with some upper-case characters are excluded. The reason is that some upper-case characters (e.g. `F` and `G`) are reserved keywords for LTL and PLTL operators, and for a more intuitive usage of the grammar it is preferred to forbid all of them instead of asking the user to remember the relatively few exceptions. Moreover, the grammar should be able to support constructs like `FGa`, i.e. no necessary spaces between operators and symbols, for better conciseness.

The reserved keywords are:

- `true`, `false`, `tt`, `ff`, the boolean constants;
- `last`, `end`, `first`, `start`, the temporal logic abbreviations;
- `F`, `G`, `H`, `M`, `O`, `R`, `S`, `U`, `V`, `W`, `X`, `Y`, the temporal operators.

```
NameStartChar ::= [a-z] | "_"
NameChar      ::= NameStartChar | [0-9]
Name          ::= NameStartChar (NameChar)*
QuotedName    ::= ('"' [^"\n\t\r]* '"') | ("'" [^'\n\t\r]* "'")
Keywords      ::= PropBooleans
                | LogicBooleans
                | "last" | "end" | "first" | "start"
                | "F" | "G" | "H" | "M" | "O" | "R" | "S" | "U" | "V"
                | "W" | "X" | "Y"
Atom          ::= (Name | QuotedName) - Keywords
```

## Boolean operators

The supported boolean operations are: negation, conjunction, disjunction, implication, equivalence and exclusion.

Follows the list of characters used for each operator:

- negation: `!`, `~`;
- conjunction: `&`, `&&`;
- disjunction: `|`, `||`;
- implication: `->`, `=>`;
- equivalence: `<->`, `<=>`;
- exclusive disjunction: `^`;

```
Non   ::= "!" | "~"
And   ::= "&" | "&&"
Or    ::= "|" | "||"
Impl  ::= "->" | "=>"
Equiv ::= "<->" | "<=>"
Xor   ::= "^"
```

## Parenthesis

We use `(` and `)` for parenthesis.

```
LeftParen  ::= "("
RightParen ::= ")"
```

## White Spaces

It is often convenient to use "white spaces" (spaces, tabs, and blank lines) to set apart the formulae for greater readability. These characters MUST be ignored when processing the text input.

# LTLf

In this section, we specify a grammar for LTLf.

## Atoms

An LTLf formula is defined over a set of *atoms*. In this context, an atom formula is defined by using the `Atom` regular language defined above:

```
LTLAtom ::= Atom
```

## Temporal operators

Here we specify the regular languages for the temporal operators.

- (Weak) Next: `X`;
- Strong Next: `X[!]`;
- (Strong) Until: `U`;
- Weak Until: `W`;
- (Weak) Release: `R`, `V`;
- Strong Release: `M`;
- Eventually: `F`;
- Always: `G`;

In EBNF format:

```
WeakNext       ::= "X"
Next           ::= "X[!]"
Until          ::= "U"
WeakUntil      ::= "W"
Release        ::= "R" | "V"
StrongRelease  ::= "M"
Eventually     ::= "F"
Always         ::= "G"
```

## Special Formulae

Special LTLf formulae are:

- `last`, meaning "the last step of the trace", semantically equivalent to `X(false)`;
- `end`, meaning "the end of the trace", semantically equivalent to `G(false)`.

In EBNF format:

```
Last ::= "last"
End  ::= "end"
```

## Grammar

```
ltl_formula ::= LTLAtom
              | PropBooleans
              | LogicBooleans
              | Last
              | End
              | LeftParen ltl_formula RightParen
              | Not ltl_formula
              | ltl_formula And ltl_formula
              | ltl_formula Or ltl_formula
              | ltl_formula Impl ltl_formula
              | ltl_formula Equiv ltl_formula
              | ltl_formula Xor ltl_formula
              | ltl_formula Until ltl_formula
              | ltl_formula WeakUntil ltl_formula
              | ltl_formula Release ltl_formula
              | ltl_formula StrongRelease ltl_formula
              | Eventually ltl_formula
              | Always ltl_formula
              | WeakNext ltl_formula
              | Next ltl_formula
```

For the semantics of these operators, we refer to [3] for the finite setting.

## Precedence and associativity of operators

The precedence and associativity of the LTL operators are described by the following table (priorities from lowest to highest). For brevity, aliases for boolean operators are omitted.

| associativity | operators |
| --- | --- |
| right | `->`, `<->` |
| left | `^` |
| left | `|` |
| left | `&` |
| right | `U`, `W`, `M`, `R` |
| right | `F`, `G` |
| right | `X`, `X[!]` |
| right | `!` |

# LDLf

In this section, we specify a grammar for LDLf.

## Temporal operators

LDLf supports two temporal operators:

- *Diamond* operator: `<regex>ldl_formula`;
- *Box* operator: `[regex]ldl_formula`;

`regex` will be presented in the next paragraph.

```
LeftDiam  ::= "<"
RightDiam ::= ">"
LeftBox   ::= "["
RightBox  ::= "]"
```

In EBNF format, an LDLf formula is defined as follows:

```
ldl_formula ::= TT
              | FF
              | LeftParen ldl_formula RightParen
              | Not ldl_formula
              | ldl_formula And ldl_formula
              | ldl_formula Or ldl_formula
              | ldl_formula Impl ldl_formula
              | ldl_formula Equiv ldl_formula
              | LeftDiam regex RightDiam ldl_formula
              | LeftBox regex RightBox ldl_formula
```

## Regular Expressions

In this section, we define the regular expression used by Diamond and Box operators.

A regular expression is defined inductively as:

- a *propositonal formula* over as set of propositional atoms.
- a *test expression*: `ldl_formula?
- a *concatenation* between two regular expressions: `regex_1 ; regex_2`
- a *union* between two regular expressions: `regex_1 + regex_2`
- a *star* operator over a regular expression: `regex*`

The symbols are listed below:

```
Test   ::= "?"
Concat ::= ";"
Union  ::= "+"
Star   ::= "*"
```

The EBNF grammar for a regular expression is:

```
propositional ::= Atom
               | True
               | False
               | LeftParen propositional RightParen
               | Not propositional
               | propositional And propositional
               | propositional Or propositional
               | propositional Impl propositional
               | propositional Equiv propositional
               | propositional Xor propositional

regex ::= propositional
        | LeftParen regex RightParen
        | regex Test
        | regex Concat regex
        | regex Union regex
        | regex Star
```

For the semantics of the operators, we refer to [3].

## Precedence and associativity of operators

The precedence and associativity of the LDL operators are described by the following table (priorities from lowest to highest). For brevity, aliases for boolean operators are omitted.

| associativity | operators |
| --- | --- |
| right | `->` , `<->` |
| left | `^` |
| left | `|` |
| left | `&` |
| N/A | `<>` , `[]` |
| left | `;` |
| left | `+` |
| left | `*` |
| left | `?` |
| right | `!` |

## PLTLf

In this section, we specify a grammar for PLTLf.

### Atoms

A PLTLf formula is defined over a set of *atoms*. In this context, an atom formula is defined by using the `Atom` regular language defined above:

```
PLTLAtom ::= Atom
```

## Temporal operators

Here we specify the regular languages for the temporal operators.

- Before: `Y`;
- Since: `S`;
- Once: `O`;
- Historically `H`

In EBNF format:

```
Before       ::= "Y"
Since        ::= "S"
Once         ::= "O"
Historically ::= "H"
```

## Special Formulae

Special PLTLf formulae are:

- `first`, meaning "the first step of the trace", semantically equivalent to `!B(true)`;
- `start`, meaning "the start of the trace", semantically equivalent to `H(false)`.

In EBNF format:

```
First ::= "first"
Start ::= "start"
```

## Grammar

```
pltl_formula ::= PLTLAtom
              | PropBooleans
              | LogicBooleans
              | First
              | Start
              | LeftParen pltl_formula RightParen
              | Not pltl_formula
              | pltl_formula And pltl_formula
              | pltl_formula Or pltl_formula
              | pltl_formula Impl pltl_formula
              | pltl_formula Equiv pltl_formula
              | pltl_formula Xor pltl_formula
              | pltl_formula Since pltl_formula
              | Once pltl_formula
              | Historically pltl_formula
              | Before pltl_formula
```

For the semantics of these operators for the finite setting, we refer to [6].

## Precedence and associativity of operators

The precedence and associativity of the LTL operators are described by the following table (priorities from lowest to highest). For brevity, aliases for boolean operators are omitted.

| associativity | operators |
| --- | --- |
| right | `->`, `<->` |
| left | `^` |
| left | `|` |
| left | `&` |
| right | `S` |
| right | `O`, `H` |
| right | `B` |
| right | `!` |

# PLDLf

In this section, we specify a grammar for PLDLf.

## Temporal operators

PLDLf supports two temporal operators:

- *Backward diamond* operator: `<<regex>>pldl_formula`;
- *Backward box* operator: `[[regex]]pldl_formula`;

`regex` is the same as defined for LDLf.

```
LeftBackwardDiam   ::= "<<"
RightBackwardDiam  ::= ">>"
LeftBackwardBox    ::= "[["
RightBackwardBox   ::= "]]"
```

In EBNF format, a PLDLf formula is defined as follows:

```
pldl_formula ::= TT
              | FF
              | LeftParen pldl_formula RightParen
              | Not pldl_formula
              | pldl_formula And pldl_formula
              | pldl_formula Or pldl_formula
              | pldl_formula Impl pldl_formula
              | pldl_formula Equiv pldl_formula
              | LeftBackwardDiam regex RightBackwardDiam pldl_formula
              | LeftBackwardBox regex RightBackwardBox pldl_formula
```

For the semantics of the operators, we refer to [6].

## Precedence and associativity of operators

The precedence and associativity of the LDL operators are described by the following table (priorities from lowest to highest). For brevity, aliases for boolean operators are omitted.

| associativity | operators |
|---|---|
| right | `->`, `<->` |
| left | `^` |
| left | `|` |
| left | `&` |
| N/A | `<<>>`, `[[]]` |
| left | `;` |
| left | `+` |
| left | `*` |
| left | `?` |
| right | `!` |

## Future work

In future versions of this standard, we would like to add:

- `Spot`-like syntactic sugars for regular expressions (SERE) and temporal operators [14,24];
- Compatibility with the PSL standard [13];

- Support full Unicode characters, so to use UTF-8 characters like ○ (U+25CB) for the Next operator and ◇ (U+25C7) for the Eventually operator etc. as alternative symbols.

---

1. You can get the sources of this document at this repository: https://github.com/marcofavorito/tl-grammars↩